# Statistics as a dynamical attractor.

Michail Zak
Senior Research Scientist (Emeritus)
Jet Propulsion Laboratory California Institute of Technology
Pasadena, CA 91109

## Abstract.

It is demonstrated that any statistics can be represented by an attractor of the solution to a corresponding system of ODE coupled with its Liouville equation. Such a non-Newtonian representation allows one to reduce foundations of statistics to better established foundations of ODE. In addition to that, evolution to the attractor reveals possible micro-mechanisms driving random events to the final distribution of the corresponding statistical law. Special attention is concentrated upon the power law and its dynamical interpretation: it is demonstrated that the underlying dynamics supports a "violent reputation" of the power-law statistics.

**1.Introduction**.

This paper presents an attempt of dynamical interpretation of statistics laws. Unlike a law in science that is expressed in the form of an analytic statement with some constants determined empirically, a law of statistics represents a function all values of which are determined empirically; in addition to that, the micro-mechanisms driving the underlying sequence of random events to their final probability distribution remained beyond any description.

The objective of this paper is to create a system of ODE that has a prescribed statistics as a dynamical attractor. Such representation allows one to reduce foundations of statistics to better established foundations of ODE, as well as to find a virtual model of micro-mechanisms that drive a sequence of random events to the final distribution of the corresponding statistical law.

It is clear that the ODE to be found cannot be based upon Newtonian dynamics since their solutions must be random. The Langeven equations must also be disqualified since their randomness is generated by random inputs. The only ODE that have intrinsically random solution are so called quantum-inspired (iQ) ODE introduced and analyzed in [1, 2, 3]. The distinguished property of these equations is that they are coupled with their Liouville equation by mean of a specially selected feedback. This feedback violates the Lipschits condition and creates instability that leads to randomness. The proposed model is represented by a modified Madelung equation [4] in which the quantum potential is replaced by different, specially chosen "computational" potential. As a result, the dynamics attains both quantum and classical properties.

Based on the Madelung version of the Schrödinger equation, the origin of randomness in quantum mechanics has been traced down to instability *generated by quantum potential* at the point of departure from a deterministic state, [4]. The instability triggered by failure of the Lipchitz condition splits the solution into a continuous set of random samples representing a hidden statistics of Schrödinger equation, i.e., the transitional stochastic process as a "bridge" to quantum world. The proposed model has similar properties, but it is not conservative, and therefore, it can have attractors.

**2. Dynamical model for simulations**.

We will start this section with a brief review of models introduced and discussed in [1,2]. For mathematical clarity, we will consider a one-dimensional motion of a unit mass under action of a force $f$ depending upon the velocity $x$ and time $t$

$$\dot{x} = f(x,t), \tag{1}$$

If initial conditions are not deterministic, and their probability density is given in the form

$$\rho_0 = \rho_0(X), \qquad where \quad \rho \geq 0, \quad and \qquad \int_{-\infty}^{\infty} \rho dX = 1 \tag{2}$$

while $\rho$ is a single- valued function, then the evolution of this density is expressed by the corresponding Liouville equation

$$\frac{\partial \rho}{\partial t} + \frac{\partial}{\partial X}(\rho f) = 0 \tag{3}$$

The solution of this equation subject to initial conditions and normalization constraints (2) determines probability density as a function of $X$ and $t$

$$\rho = \rho(X,t) \tag{4}$$

Let us now specify the force $f$ as a feedback from the Liouville equation

$$f(x,t) = \varphi[\rho(x,t)] \tag{5}$$

and analyze the motion after substituting the force (5) into Eq.(1)

$$\dot{x} = \varphi[\rho(x,t)], \tag{6}$$

*Remark*. Theory of stochastic differential equations makes distinction between the random *variable x(t)* and its *values X* in probability space.

This is a fundamental step in our approach. Although theory of ODE does not impose any restrictions upon the force as a function of space coordinates, the Newtonian physics does: equations of motion are never coupled with the corresponding Liouville equation. Moreover, as shown in [1, 2], such a coupling leads to non-Newtonian properties of the underlying model. Indeed, substituting the force $f$ from Eq. (5) into Eq. (3), one arrives at the *nonlinear* and, in general*, non-reversible* equation for evolution of the probability density

$$\frac{\partial \rho}{\partial t} + \frac{\partial}{\partial X}\{\rho\varphi[\rho(X,t)]\} = 0 \tag{7}$$

Now we will demonstrate the destabilizing effect of the feedback (6). For that purpose, it should be noted that the derivative $\partial\rho/\partial x$ must change its sign, at least once, within the interval $-\infty < x < \infty$, in order to satisfy the normalization constraint (2). But since

$$Sign\frac{\partial \dot{x}}{\partial x} = Sign\frac{d\varphi}{d\rho}\,Sign\frac{\partial \rho}{\partial x} \tag{8}$$

there will be regions of $x$ where the motion is unstable, and this instability generates randomness with the probability distribution guided by the Liouville equation (7).

**3. Statistics as a dynamical attractor.**

Let us consider Eqs. (1) and (7) defining *f* as the following function of probability density

$$f = \frac{1}{\rho(x,t)}\int_{-\infty}^{x}[\rho(\zeta,t) - \rho^*(\zeta)]d\zeta \tag{9}$$

With the feedback (9), Eqs. (1) and (7) take the form, respectively

$$\dot{x} = \frac{1}{\rho(x,t)}\int_{-\infty}^{x}[\rho(\zeta,t) - \rho^*(\zeta)]d\zeta \tag{10}$$

$$\frac{\partial \rho}{\partial t} + \rho(t) - \rho^* = 0 \tag{11}$$

The last equation has the analytical solution

$$\rho = (\rho_0 - \rho^*)e^{-t} + \rho^* \tag{12}$$

Subject to the initial condition

$$\rho(t=0) = \rho_0 \tag{13}$$

this solution converges to a prescribed, or target, stationary distribution $\rho^*(x)$. Obviously the normalization condition for $\rho$ is satisfied if it is satisfied for $\rho_0$ and $\rho*$.

Substituting the solution (12) in to Eq. (10), one arrives at the ODE that simulates the stochastic process with the probability distribution (12)

$$\dot{x} = \frac{e^{-t}}{[\rho_0(x) - \rho^*(x)]e^{-t} + \rho^*(x)}\int_{-\infty}^{x}[\rho_0(\zeta) - \rho^*(\zeta)]\ d\zeta \tag{14}$$

As notices above, the randomness of the solution to Eq. (14) is caused by instability that is controlled by the corresponding Liouville equation. It should be emphasized that in order to run the stochastic process started with the initial distribution $\rho_0$ and approaching a stationary process with the distribution $\rho^*$, one should substitute into Eq. (14) *analytical expressions* for these functions.

It is reasonable to assume that the solution (12) starts with sharp initial condition

$$\rho_0(X) = \delta(X) \quad (15)$$

As a result of that assumption, all the randomness is supposed to be generated *only* by the controlled instability of Eq. (14). Substitution of Eq. (15) into Eq. (14) leads to two different domains of $x$: $x \neq 0$ and $x \equiv 0$. The solution for the first domain is

$$\int_{-\infty}^{x} \rho^*(\zeta) d\zeta = \frac{C_1}{e^{-t} - 1}, \qquad x \neq 0 \quad (16)$$

Indeed, $\dot{x} = \frac{e^{-t}}{[\rho_0(x) - \rho^*(x)]e^{-t} + \rho^*(x)} \int_{-\infty}^{x} [\rho_0(\zeta) - \rho^*(\zeta)]\ d\zeta = \frac{e^{-t}}{\rho^*(x)(e^{-t} - 1)} \int_{-\infty}^{x} \rho^*(\zeta) d\zeta$

whence $\frac{\rho^*(x)}{\int_{-\infty}^{x} \rho^*(\zeta) d\zeta} dx = \frac{e^{-t}}{e^{-t} - 1} dt$ . Therefore, $\ln \int_{-\infty}^{x} \rho^*(\zeta) d\zeta = \ln \frac{C}{e^{-t} - 1}$ and that leads to Eq. (16).

The solution for the second domain is

$$x \equiv 0 \quad (17)$$

Eq. (17) represents a singular solution, while Eq. (16) is a regular solution that includes arbitrary constant $C$. The regular solutions are unstable at $t=0$, $|x| \to 0$ where the Lipschitz condition is violated

$$\frac{d\dot{x}}{dx} \to \infty \qquad at \qquad t \to 0 \quad ,|x| \to 0 \quad (18)$$

and therefore, an initial error always grows generating *randomness*.

Let us analyze the behavior of the solution (16) in more details. As follows from this solution, all the particular solutions intersect at the same point $x$=0 at $t$=0, and that leads to non-uniqueness of the solution due to violation of the Lipcshitz condition. Therefore, the same initial condition $x$=0 at $t$=0 yields infinite number of different solutions forming a family (16); each solution of this family appears with a certain probability guided by the corresponding Liouville equation (11). For instance, in cases plotted in Fig.1 the "winner" solution is

$$x_1 = \varepsilon \to 0, \qquad \rho(x_1) = \rho_{\max} \quad (19)$$

since it passes through the maximum of the probability density (11). However, with lower probabilities, other solutions of the family (13) can appear as well. Obviously, this is a non-classical effect. Qualitatively, this property is similar to those of quantum mechanics: the system keeps all the solutions simultaneously and displays each of them "by a chance", while that chance is controlled by the evolution of probability density (11). It should be emphasized that in the ideal case, when no noise is present, the choice of displaying a certain solution is made by the system only once, at $t$=0, i.e. when it departs from a deterministic to a random state; since than, it stays with this solution as long as the Liouville feedback is present. However, strictly speaking, an actual realization of the trajectory may be affected by a non-Lipschitz-originated instability at $t=0$; as a result, small initial errors may grow exponentially, and the motion will be randomly deviated from the theoretical trajectory in such a way that a moving particle visits all the possible trajectories with the probability prescribed by the Liouville equation.

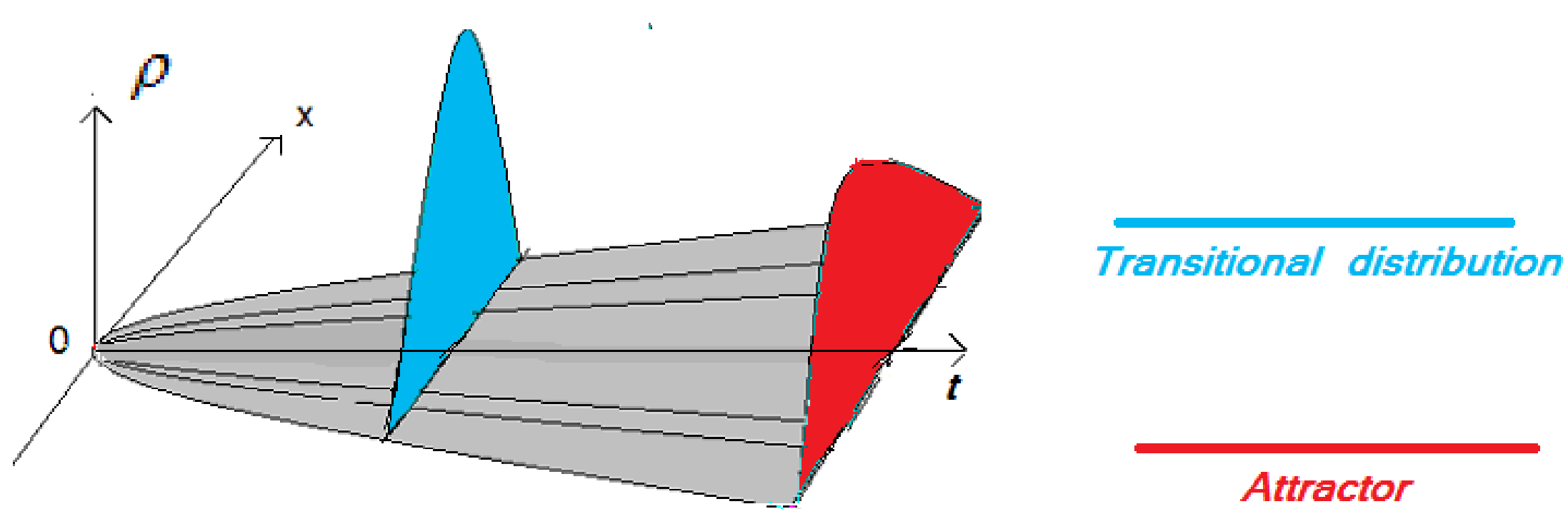

**Figure 1, Statistics as a dynamical attractor.**

The model is generalized to $n$-dimensional case simply by replacing $x$ with a vector $x = x_1, x_2, ... x_n$ since Eq. (11) does not include space derivatives.

It can be written as

$$\dot{x}_i = \frac{1}{n\rho(x_{1,}...x_n, t)} \int_{-\infty}^{x_i} [\rho(\zeta_i, x_{j\neq i}, t) - \rho^*(\zeta_i, x_{j\neq i})] d\zeta_i \qquad \text{i=1, 2…n} \qquad (20)$$

where

$$\rho = (\rho_0 - \rho^*)e^{-t} + \rho^* \qquad (21)$$

There has been demonstrated that the "random" dynamics considered above represents a fundamental departure from both Newtonian and statistical mechanics, and although it has some quantum-like properties,[1,3], it is rather a quantum-classical hybrid that links to model of Livings introduced in [ 1]. Nevertheless the model is fully consistent with the theory of differential equations and stochastic processes.

**4.Examples**.

Prior to analysis of the proposed approach, we consider some instructive examples.

Let us start with the following normal distribution

$$\rho^*(X) = \frac{1}{\sqrt{2\pi}} e^{-\frac{X^2}{2}} \qquad (22)$$

Substituting the expression (22) and (15) into Eq. (16) at $X$=$x$, one obtains

$$x = erf^{-1}(\frac{C}{e^{-t}-1}), \quad v \neq 0 \qquad (23)$$

The dynamics driving random events to normal distribution is plotted in Fig.4.

Let us now choose the final density as a uniform distribution

$$\rho^*(X) = \{\frac{1/(b-a) \quad if \quad -a \leq X \leq b}{0 \quad otherwise}, \qquad a > 0 \qquad (24)$$

Then

$$\int_{-\infty}^{x} \rho^*(\zeta) d\zeta = \{\frac{(v-a)/(b-a) \quad if \quad a \leq x < b}{1 \quad if \quad x \geq b} \qquad (25)$$

Substituting Eq. (25) into Eq. (16) at $X$=$x$ >0 yields

$$x = a + \frac{C}{e^{-t}-1}, \qquad 0 > C > a-b, \qquad a \leq x \leq b, \ t \geq \ln\frac{b-a}{C+b-a} \qquad (26)$$

As another example, let us choose the target density $\rho^*$ as the Student's distribution, or so called power law distribution

$$\rho^*(X) = \frac{\Gamma(\frac{\nu+1}{2})}{\sqrt{\nu\pi}\Gamma(\frac{\nu}{2})}(1+\frac{X^2}{\nu})^{-(x+1)/2} \qquad (27)$$

Substituting the expression (24) into Eq. (16) at $X$=$x$, and ν=1, one obtains

$$x = \cot(\frac{C}{e^{-t}-1}) \qquad for \qquad x \neq 0 \qquad (28)$$

The evolution of a particular solutions at C=1, 10, and 100 is plotted in Fig 2. Here $C$ is the parameter representing a particular random sample of the solution. This solution describes (in an implicit form) a one-parametric family of dynamical processes $x = x(t, C)$. Each scenario (for a fixed $C$) occurs with the current probability (21) that asymptotically approaches the power-law distribution (27) at critical points. Those critical points that occur at large $x$ can be associated with catastrophes.

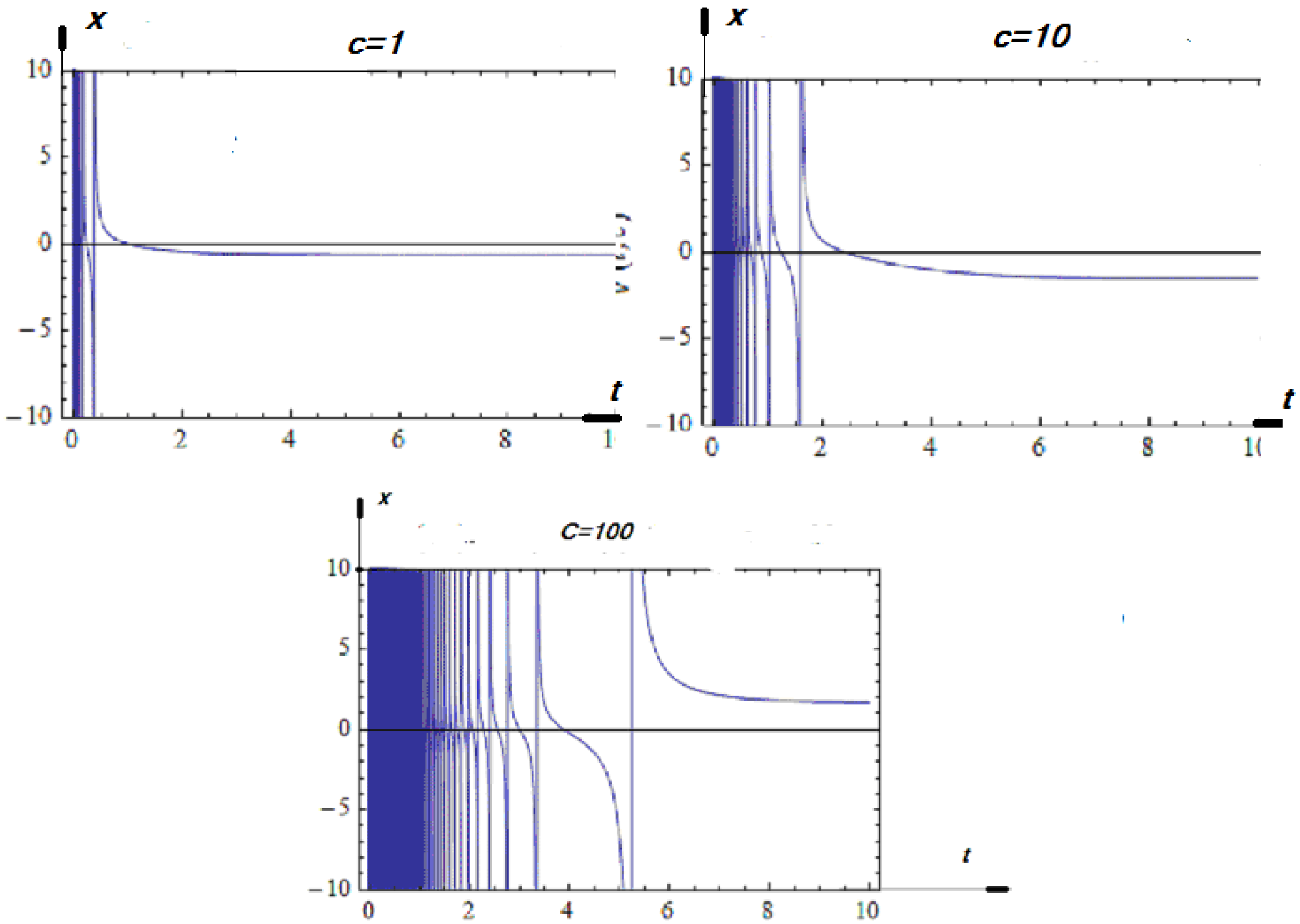

**Figure 2. The evolution of a particular solutions at C=1, 10, 100.**

**5. A mystery of power-law statistics.**

The objective of this section was inspired by a mysterious power-law statistics that predicts social catastrophes: wars, terrorist attacks, market crushes est. Resent interest in literature is concentrated on half-a-century finding that the severity of interstate wars is power-law distributed, and that belongs to the most striking empirical regularities in world politics. Surprisingly, similar catastrophes were identified in physics (Ising systems, avalanches, earthquakes), and even in geometry (percolation). Although all these catastrophes have different origins, their similarity is based upon the power law statistics, and as a consequence, on scale invariance, self-similarity and fractal dimensionality, [5]. According to the theory of self-organized criticality, that explains the origin of this kind of catastrophes, each underlying dynamical system is attracted to a critical point separating two qualitatively different states (phases). This attraction is represented by a relaxation process of slowly driven system. Transitions from one phase to another are accompanied by sudden release of energy that can be associated with a catastrophe, and the severity of the catastrophe is power law distributed. However, in order to overcome the critical point and enter a new phase, a slow input of *external* energy is required. The origin of this energy is well understood in physical systems, but not in social ones, since there are no well established models of social dynamics. For that reason, we turn to the previous section and start with comparison the underlying dynamics of normal and power law distribution, (see Figs. 3.4, and 5). Let us recall that the normal distribution is commonly encountered in practice, and is used throughout statistics, natural sciences, and social sciences as a simple model for complex phenomena. For example, the observational error in an experiment is usually assumed to follow a normal distribution, and the propagation of uncertainty is computed using this assumption. But statistical inference using a normal distribution is not robust to the presence of outliers (data that is unexpectedly far from the mean, due to exceptional circumstances, observational error, etc.). When outliers are expected, data may be better described using a heavy-tailed distribution such as the power-law distribution. As demonstrated in Fig. 3, normal and power law distributions have very close configurations excluding the tails. However despite of that, the types of the random events described by these statistics are of fundamental difference. Indeed, processes described by normal distributions are usually coming from physics, chemistry, biology, est., and they are characterized by a smooth evolution of underlying dynamical events. On the contrary, the processes described by power laws are originated from events driven by human decisions (wars, terrorist acts, market crushes), and therefore, they are associated with catastrophes. Surprisingly, the 3D plots of Eqs.(23) and (28) (see Figs.4 and 5) describing dynamics that drives random events to the normal and the power law distributions, respectively, demonstrate the same striking difference between these distributions, that is: a smooth evolution to normal distribution, and "violent", full of densely distributed discontinuities (see Fig. 2) transition to power law distribution. In Fig.2, *C* is the parameter representing a particular random sample of the solution. This solution describes (in an implicit form) a one-parametric family of dynamical processes $x = x(t, C)$. Each scenario (for a fixed *C*) occurs with the current probability (21) that asymptotically approaches the power-law distribution (27) at critical points. Those critical points that occur at large *x* can be associated with catastrophes.
Is this a coincidence? Indeed, the proposed random dynamics is based upon global assumptions, and it does not bear any specific information about a particular statistics as an attractor. However the last statement should be slightly modified:

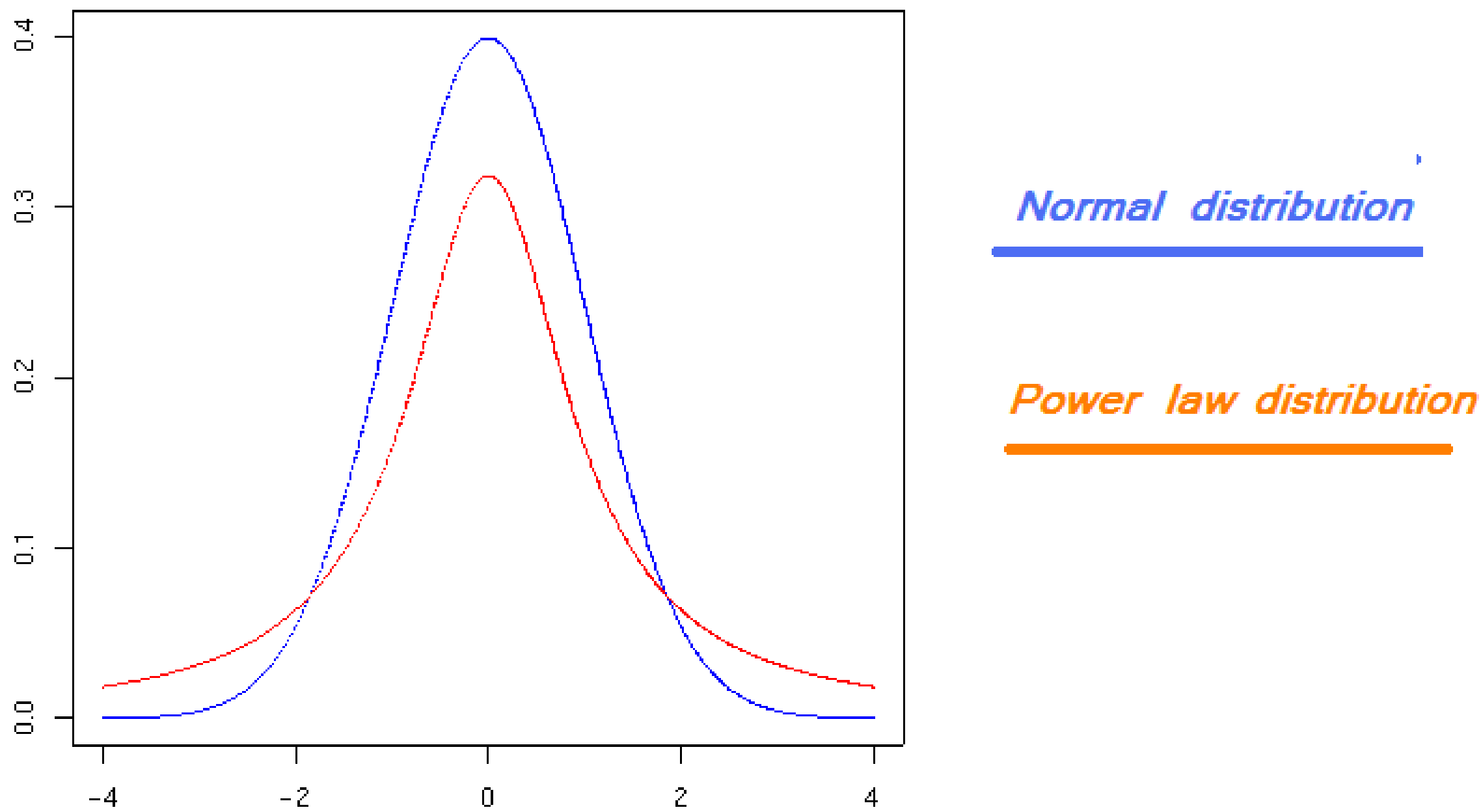

**Figure 3. Normal and power law distributions.**

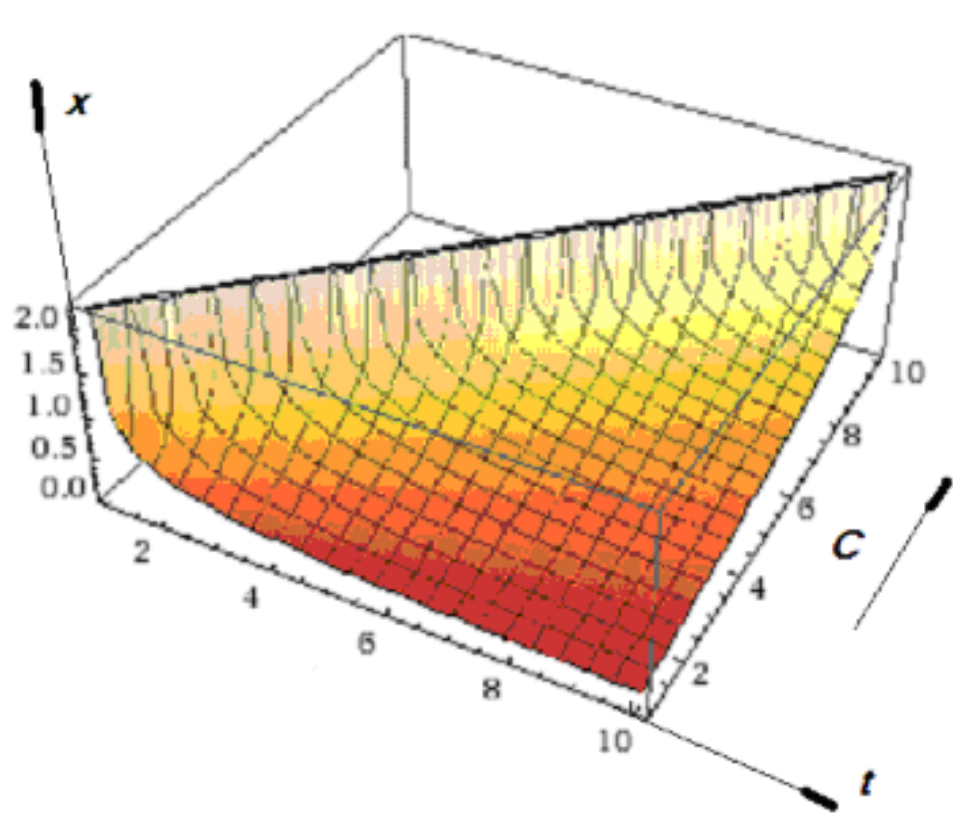

**Figure 4. Dynamics driving random events to normal distribution.**

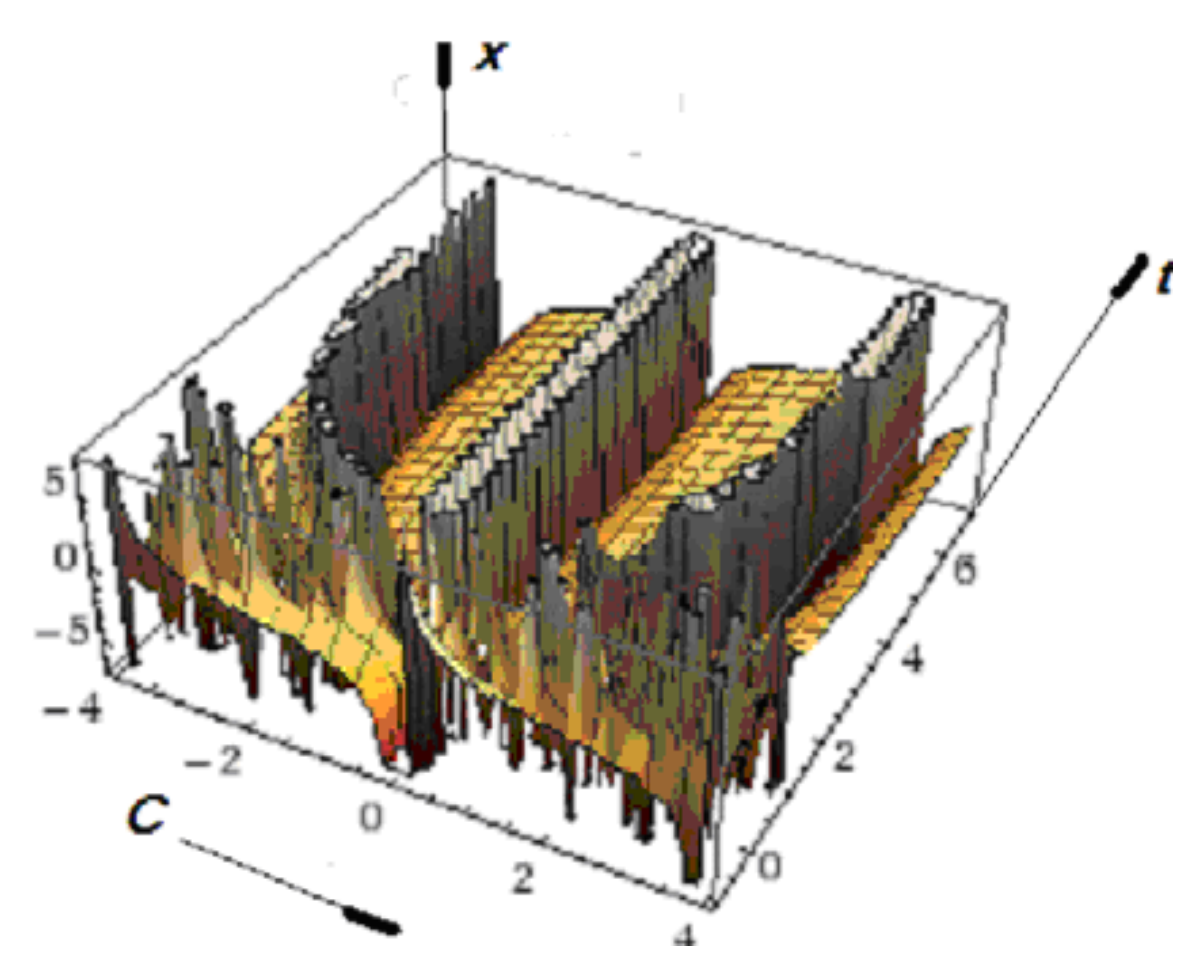

**Figure 5. Dynamics driving random events to power law distribution.**

actually the model of random dynamics is tailored to describe Livings' behavior, and in particular, decision making process,[1.3 ]. Is that why the random dynamics captures "violent" properties of power law statistics that is associated with human touch? We will discuss possible answers to this question below.
First of all we have to consider the problem of uniqueness of a dynamical system which solutions approach a preset attractor. In classical dynamics the answer to this problem is clear: there is infinite number of different dynamical systems that approach the same attractor, and that is true for static, periodic and chaotic attractors. Although in random dynamics we are dealing with a stochastic attractor that represents a preset statistics, the answer is similar. Indeed, let us turn to Eq. (10) and add an arbitrary terms as following

$$\dot{x} = \frac{1}{\rho(x,t)}\{\int_{-\infty}^{x}[\rho(\zeta,t) - \rho^*(\zeta)]d\zeta + A + B(t)\} \quad (29)$$

where $A$ is an arbitrary constant, and $B(t)$ is arbitrary function of time.
It easily verifiable that this modification does not change Eq.(11),

$$\frac{\partial \rho}{\partial t} + \rho(t) - \rho^* = 0 \quad (30)$$

and therefore, the attractor $\rho^*$ remains the same, despite of different solution of Eq.(29).
That excludes the possibility of a rigorous proof that the random dynamics necessarily describes the same random events that constitute the corresponding statistics. But in order to explain, at least, some correlations between them, we will turn to a less rigorous approach known as an Occam Razor. After several modifications, a "scientific" version of this approach can be formulated as following: one should proceed to simpler theories until simplicity can be traded for greater explanatory power. Obviously application of this principle does not guarantee a success, however, there is known some encouraging results of its applications: In science, Occam's razor is used as a heuristic (rule of thumb) to guide scientists in the development of theoretical models rather than as an arbiter between published models. In physics, parsimony was an important heuristic in the formulation of special relativity by Albert Einstein, the development and application of the principle of least action by Pierre Louis Maupertuis and Leonhard Euler, and the development of quantum mechanics by Ludwig Boltzmann, Max Planck, Werner Heisenberg and Louis de Broglie. There have been attempts to derive Occam's Razor from probability theory, notable attempts made by Harold Jeffreys and E. T. Jaynes. Using Bayesian reasoning, a simple theory is preferred to a complicated one because of a higher prior probability. Hence, application of the Razor requires a definition of the concept of complexity/simplicity. However in our case the simplest dynamical system is obvious: it follows from Eq. (29) when

$$A \equiv 0, \qquad B \equiv 0 \quad (31)$$

i.e. the ***most likely*** scenario describing dynamics that drives random events to power law distribution is still a dynamical system Eq.(10) and (11) that was discussed above. However if additional information about the random events dynamics is available, it should be incorporated to the enlarged model Eq. (29) through a best-fit adjustment of the constant $A$ and the parameterized function $B(t)$ thereby trading simplicity for greater explanatory power. We have to emphasize that the enlarged dynamical system still has the same power-law statistics as an attractor.

**6. General case.**

Based upon the proposed model, a simple algorithm for finding a dynamical system that is attracted to a preset $n$-dimensional statistics can be formulated. The idea of the proposed algorithm is very simple: based upon the system (20), and (21), introduce the probability density $\rho^*(x_1, x_2, ... x_n)$ representing the statistics to which the solution of Eq. (21) is to be attracted, and insert it in Eqs. (20) and (21). The solution of Eq. (20) will eventually approach a stochastic attractor which probability density coincides with the preset statistics. As in the case of the one-dimensional power-law statistics, here the solution of Eq. (29) will present the ***most likely*** scenario describing dynamics that drives random events to preset statistics. Moreover if additional information about the random events dynamics is available, it should be incorporated to the enlarged model of Eq. (20) that is

$$\dot{v}_i = \frac{1}{n\rho(v_1, ... v_n, t)}[\int_{-\infty}^{v_i}[\rho(\zeta_i, v_{j\neq i}, t) - \rho^*(\zeta_i, v_{j\neq i})]d\zeta_i + A_i + B_i(t) + \sum_{j\neq i}^{n} T_{ij}\tanh v_j] \quad (32)$$

through a best-fit adjustment of the constants $A_i$, the parameterized functions $B_i(t)$), as well as the constants $T_{ij}$ that introduce zero-divergence terms for n>1, thereby trading simplicity for greater explanatory power and without a change of the preset statistics as is the one-dimensional case.

**Remark**. It is easily verifiable that the augmented terms in Eqs.(32) do not effect the corresponding Liouville equation (30), and therefore, they do not change the static attractor in the probability space (that corresponds to the stochastic attractor in physical space).. However, they may significantly change the configuration of the random trajectories in physical space making the dynamics more sophisticated.

It should be noticed that the proposed approach imposes a weak restriction upon the *space structure* of the function $\rho(\{x\})$ : it should be only integrable since there is no space derivatives included in Eq. (30). This means that $\rho(\{x\})$ is not necessarily to be differentiable. For instance, it can be represented by a Weierstrass-like function $f(x) = \sum_0^\infty a^n \cos(b^n \pi x)$, where $0 < a < 1$, $b$ is a positive odd integer, and $ab > 1 + 1.5\pi$ .

**7. Discussion and conclusion.**

The main result being proved in this paper is that ***for any statistics it can be found a set of dynamical systems having a stochastic attractor whose probability density is identical to the underlying statistics, and therefore, the statistics becomes a static attractor in probability space.*** It has been hypothesized that the simplest system in each set is likely to present a first approximation to the scenario simulating dynamics that drives random events to the corresponding stochastic attractor. It was demonstrated how to find these dynamical systems given the underlying statistics. Special attention was concentrated on power-law statistics, and its interpretation, with help of the dynamical system, is proposed.

Let us discuss now possible contribution of the presented result to the foundation of statistics and probability in general. There are two broad categories of probability interpretations which can be called "physical" and "evidential" probabilities.

Physical probabilities, which are also called objective or frequency probabilities, are associated with random physical systems such as roulette wheels, rolling dice and radioactive atoms. In such systems, a given type of event (such as the dice yielding a six) tends to occur at a persistent rate, or "relative frequency", in a long run of trials. Physical probabilities either explain, or are invoked to explain, these stable frequencies. Thus talk about physical probability makes sense only when dealing with well defined random experiments.

Evidential probability, also called Bayesian probability (or subjectivist probability), can be assigned to any statement whatsoever, even when no random process is involved, as a way to represent its subjective plausibility, or the degree to which the statement is supported by the available evidence. On most accounts, evidential probabilities are considered to be degrees of belief, defined in terms of dispositions to gamble at certain odds.

Comparing these alternative interpretations in view of the random dynamics approach introduced above, one concludes that the latter partially unites these two extremes. Indeed, the randomness there is of dynamical origin: it is generated by the Liouville equation along with non- Lipcshitz instability caused by the feedback to the equations of motion. However the dynamical origin does not mean a physical one since the random dynamics is not Newtonian and not Quantum: it is rather a quantum-classical hybrid that describes behavior of Livings, [1,2,3]. That is why this dynamics includes “human factor”: self-image, self-awareness, collective mind, est., and that place it in between of physical and Bayesian interpretations. But in addition to that, the dynamical interpretation introduced above has a solid mathematical foundation in the form of ordinary differential equations.

**References.**

1. Zak, M., 2007, Physics of Life from First Principles, EJTP **4**, No. 16(II), 11–96

2. Zak, M., 2008, Quantum-inspired maximizer, JOURNAL OF MATHEMATICAL PHYSICS **49**, 042702

3. Zak, M., 2010, Introduction to quantum-inspired intelligence, PHYSICS ESSAYS **23**, 1

4. Zak, M., 2009, Hidden statistics of Schrödinger equation, PHYSICS ESSAYS **22**, 2

**5.** Christensen, K., Moloney, N., 2005, Complexity and Criticality**,** Imperial College Press